\documentclass[10pt]{iopart}
\usepackage{iopams}
\usepackage[colorlinks=true,urlcolor=blue,citecolor=blue,linkcolor=blue]{hyperref}
\usepackage{graphicx}

\begin{document}

\title[Classical evolution in quantum systems]{Classical evolution in quantum systems}
\author{J Sperling$^{1}$ and I A Walmsley$^{2,3}$}
\address{$^1$ Integrated Quantum Optics Group, Applied Physics, Paderborn University, 33098 Paderborn, Germany}
\address{$^2$ Blackett Laboratory, Imperial College London, London SW7 2BW, United Kingdom}
\address{$^3$ Clarendon Laboratory, University of Oxford, Parks Road, Oxford OX1 3PU, United Kingdom}
\ead{jan.sperling@upb.de}

\begin{abstract}
	We investigate quantum effects in the evolution of general systems.
	For studying such temporal quantum phenomena, it is paramount to have a rigorous concept and profound understanding of the classical dynamics in such a system in the first place.
	For this reason, we derive from first principles equations of motions that describe the classical propagation in quantum systems.
	A comparison of this classical model with the actual temporal quantum behavior enables us to identify quantum phenomena in the system's dynamics and distinguish them from static quantum features at individual points in time.
	For instance, we show how Newton's second law emerges as a special case of our general treatment, connecting it to a Schr\"odinger-type equation.
	As applications of our universal technique, we analyze nonlinear optical processes, semiclassical models, and the multipartite entanglement dynamics of macroscopic ensembles.
\end{abstract}

\date{\today}
\submitto{\PS}


\maketitle
\ioptwocol

\section{Introduction}

	Quantum physics is a cornerstone of modern science, and the laws governing the quantum domain provide remarkable insights into fundamental principles of nature.
	For example, interference patterns observed in a double-slit experiment are inconsistent with the time propagation of classical particles;
	however, this and other phenomena are easily explained on the basis of the Schr\"odinger equation \cite{S26a,S26b}.
	In contrast, other interference effects are only superficially related to quantum properties, such as interference resulting from classical optics \cite{MW95}.
	Therefore, to decide whether a process is genuinely quantum, the classical counterpart to a quantum evolution is required.

	Moreover, quantum features are essential for the development of quantum technologies \cite{NC00,DM03}.
	New applications exploit the resources provided by quantum systems to perform tasks not achievable by classical means, \textit{e.g.} quantum teleportation \cite{BBCJPW93} and dense coding \cite{BW92}.
	Other protocols rely on quantum properties to improve the communication security, such as quantum key distribution \cite{BB84,GRTZ02}.
	Thus, the classical limitations of the underlying processes have to be known in order to assess possible benefits of utilizing quantum phenomena as a resource in applications.

	An early approach to certifying that a quantum evolution is incompatible with classical dynamics is the Leggett-Garg inequality \cite{LG85}; see \cite{ELN14} for a review.
	Similarly to the concept of nonlocallity, the violation of this and related inequalities enables us to uncover temporal quantum correlations \cite{BE14,BKMPP15}.
	The question of causality in quantum physics is another closely related concept, which also addresses the temporal quantumness of a process \cite{ABCFGB15,B16,RMSR17,CGB17}.
	Furthermore, noncommutative features in the mathematical formulation of quantum physics lead to quantum effects based on time-ordering \cite{QS14,QS15}.
	For example, the resulting nonclassical propagation of light yields multi-time quantum correlations for nonlinear optical processes \cite{KSV16,KVS17}.

	Despite those advances and the general importance of temporal correlations, most frequently, time-independent quantum phenomena are studied.
	Nevertheless, investigations of time-independent effects already led to remarkable insights into the quantum characteristics of systems when compared to their classical counterparts.
	In particular, the concept of quantum coherence characterizes the quantum nature in terms of quantum superpositions \cite{LM14,BCP14,SV15,MS16,WY16}; see \cite{SAP17} for a review and \cite{RBCLLAWFP18} for a recent in-depth analysis.
	In this framework, different families of classical states can be considered for different physical scenarios \cite{P72,MV96,MMSZ97}.
	Then, quantum coherence is defined as the inability to describe the state of a system in terms of such classical reference states and classical statistics.
	Moreover, this notion of quantumness includes quantum entanglement between different degrees of freedom \cite{SAWV17,SLR17,SPBW17}, which is arguably the most prominent and versatile quantum phenomena and resource for quantum protocols \cite{HHHH09}.

	In a first attempt to characterize the coherent quantum evolution, a process map can be considered that maps an initial state to a state at a later time.
	In this input-output formalism, the quantumness is characterized via the potential of the underlying map to create quantum coherence from a classical input state \cite{NDDGMOBHH03,BF05,SKVGZB13,SV15}.
	This renders it possible, for example, to investigate quantum correlations between a system and a bath \cite{H17,CGLCN18} after a given time and to study the temporal propagation of entanglement \cite{KMTKAB08,TMB08,I09,GG12}.
	Still, such methods are limited as they do not address the quantum nature of the continuous dynamics itself.
	Rather, the generation and degradation of quantum coherence are quantified separately after each time step, and the continuum of intermediate points in time is not considered.

	To overcome this restriction, we recently introduced a method to specifically analyze the entanglement dynamics \cite{SW17a}.
	This was achieved by constraining quantum trajectories to nonentangled (\textit{i.e.} separable) ones.
	However, this approach does not encompass other forms of temporal quantum phenomena based on more general notions of quantum coherence.

	In this contribution, we formulate a method to determine the classical evolution in a quantum system for studying general quantum coherence phenomena over time.
	Our technique is based on the principle of least action, by applying it to the restricted domain of a given set of classical states.
	This results in equations of motion whose solution is confined to this classical domain and can be compared with the quantum propagation to directly uncover the quantum properties of the dynamics.
	The general applicability of our approach is demonstrated with several examples to assess the time-dependent quantum behavior in different systems, such as nonlinear processes in optics and the evolution of entanglement in highly multipartite systems.
	Moreover, our technique bridges the gap between classical and quantum systems as it, among other examples, enables us to model semiclassical and even Newtonian dynamics consistently in terms of generalized Schr\"odinger-type equations.

	In summary, we devise a versatile method which enables us to describe a classical evolution in quantum systems for characterizing time-dependent quantum phenomena.
	This novel approach enables us to tell static and dynamic quantum features in large composite systems apart.
	Furthermore, the methodology devised here provides a consistent framework to study time-dependent properties of the quantum-classical transition.

	The paper is structured as follows.
	Preliminary considerations are provided in section \ref{sec:Preliminaries}.
	The main derivation of the equations of motions for pure classical states is then performed in section \ref{sec:EOM} and complemented with discussions and an extension to mixed states, sections \ref{sec:Properties} and \ref{sec:Mixed}.
	Several applications are studied in section \ref{sec:Applications}, covering the evolution of pure and mixed states in single and composite systems.
	We conclude in section \ref{sec:Conclusion}.

\section{Preliminaries and motivation}\label{sec:Preliminaries}

	For formulating self-consistent models of the classical evolution in quantum systems, let us recapitulate two fundamental ingredients.
	Firstly, the notion of quantum coherence is revisited.
	Then, the general approach for theoretically describing the dynamics of arbitrary systems is briefly reviewed.
	Thirdly, we further motivate our treatment by analyzing flaws in the predominantly applied input-output formalism.

\subsection{Quantum coherence}

	The theory of quantum coherence offers different notions of classical states, addressing different physical models \cite{SAP17}.
	Let us consider an abstract family of such predefined classical states, $|\psi(\boldsymbol q)\rangle$, parametrized by an, in general, complex vector $\boldsymbol q=(q_0,q_1,q_2,\ldots)$.
	A mixed classical state $\hat\rho$---also referred to as an incoherent state---is a statistical ensemble of the considered selection of pure reference states,
	\begin{equation}
		\label{eq:Incoherent}
		\hat\rho=\int \rmd P(\boldsymbol q)|\psi(\boldsymbol q)\rangle\langle\psi(\boldsymbol q)|,
	\end{equation}
	where $P$ is a probability distribution over the parameter space (likewise, configuration space) for $\boldsymbol q$.
	Whenever a state cannot be written in this form, this state exhibits quantum coherence which is inconsistent with the classical model under study.

	Quantum coherence applies to a wide range of physical systems.
	Since our method should meet this broad applicability, we focus on this general notion of classicality first and study specific examples later.
	For instance, see section \ref{sec:optics} for classical wave-like states forming the classical reference in optics, section \ref{sec:semiclassical} for semiclassical models of classical-quantum interactions, and section \ref{sec:entanglement} for separable states as the reference for classical correlations.

	Furthermore, the notion of quantum coherence can be extended to multipartite systems using tensor-product states \cite{SAP17}, $|\psi^{(1)}(\boldsymbol q^{(1)})\rangle\otimes|\psi^{(2)}(\boldsymbol q^{(2)})\rangle\otimes\cdots$, where $|\psi^{(j)}(\boldsymbol q^{(j)})\rangle$ are classical states in the first ($j=1$), second ($j=2$), etc. subsystem.
	The extension to mixed incoherent states is given by statistical mixtures of such pure classical states as done for the single-partite case in \eref{eq:Incoherent}.
	The notion of quantum coherence then includes single- and multipartite quantum superpositions, including entanglement; see \cite{SAWV17} for an in-depth analysis.

\subsection{Dynamics}

	The second key point for the following analysis concerns the general treatment of the propagation in time of general physical systems.
	In mathematical physics, the evolution---including arbitrary constraints---can be described in the framework of Lagrangian mechanics.
	For this reason, we may consider our set of generalized coordinates $\boldsymbol q$ and a Lagrangian $\mathcal L$ that depends on $\boldsymbol q$ and $\dot{\boldsymbol q}=\rmd {\boldsymbol q}/\rmd t$ for a closed system.
	For instance, the classical and quantum dynamics of particles and fields can be formulated in this manner; see the additional introductions to the related topics of field quantization and calculus of variations in \cite{GR96,D12}.

	The quantity action, defined as $\mathcal S=\int \rmd t\,\mathcal L$, then determines the evolution of the system.
	That is, the least action principle---a vanishing variation of the action, $\delta\mathcal S=0$---yields the Euler-Lagrange equations,
	\begin{eqnarray}
		\label{eq:EulerLagrange}
		\frac{\rmd}{\rmd t}\frac{\partial \mathcal L}{\partial \dot{\boldsymbol q}^{\ast}}=\frac{\partial \mathcal L}{\partial {\boldsymbol q}^{\ast}},
	\end{eqnarray}
	to describe the dynamics, where the conjugation can be dropped for real-valued parameters ${\boldsymbol q}$.

	In quantum physics in particular, the Lagrangian can be given as
	\begin{equation}
		\label{eq:Lagrangian}
		\mathcal L=\frac{\rmi\hbar}{2}\left(
			\langle\Psi|\dot\Psi\rangle-\langle\dot\Psi|\Psi\rangle
		\right)
		-\langle\Psi|\hat H|\Psi\rangle
	\end{equation}
	where $\hat H$ is the Hamilton operator.
	For example, the Euler-Lagrange equations for $\boldsymbol q=|\Psi\rangle$ directly yield the Schr\"odinger equation \cite{GR96},
	\begin{eqnarray}
		\label{eq:Schroedinger}
		\rmi\hbar|\dot\Psi\rangle &=& \hat H|\Psi\rangle,
	\end{eqnarray}
	which is further discussed in connection to our method in more details later (section \ref{sec:Sanity}).

\subsection{Additional motivation}

\begin{figure}
	\includegraphics[width=.99\columnwidth]{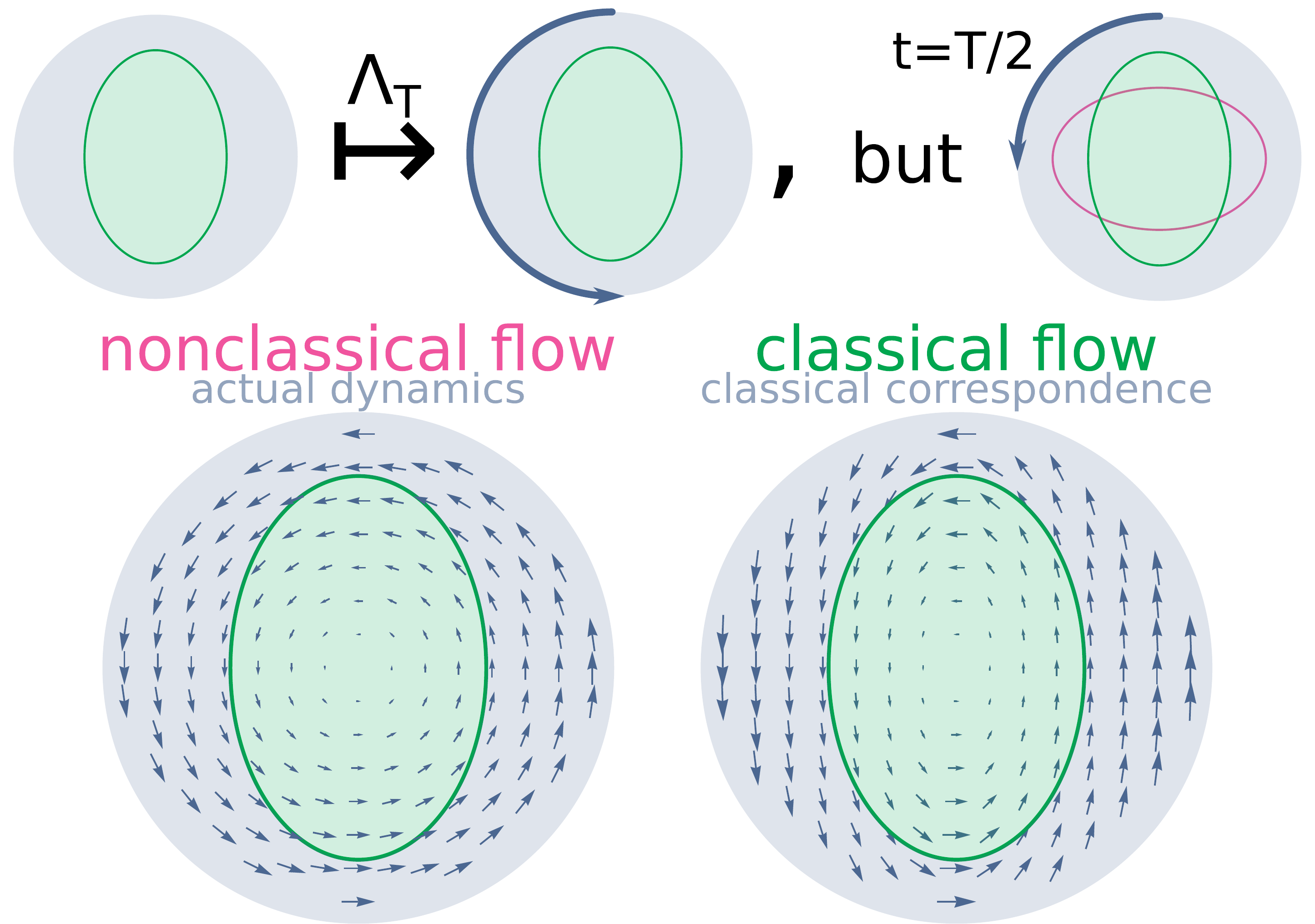}
	\caption{
		Conceptual idea of the quantum (\textit{i.e.} nonclassical) and classical dynamics.
		The circles includes all physical states; the embedded, vertically elongated ellipses represent classical states and incoherent mixtures thereof.
		A process is exemplified by a rotation.
		The top plots show that even if the process $\Lambda_T$ is classical after a $180^\circ$ rotation, intermediate times can overlap with the nonclassical domain, \textit{e.g.} a $90^\circ$ rotation.
		The bottom-left plot depicts the infinitesimal rotation (likewise, flow), which can take classical states to nonclassical ones.
		Restricting this flow to the classical domain, bottom-right plot, we obtain an infinitesimal evolution which is consistent with the set of classical states and most closely resembles the actual dynamics.
	}\label{fig:Outline}
\end{figure}

	The common definition of a classical process is given in terms of the process map, $\Lambda_T:\hat\rho\mapsto\Lambda_T(\hat\rho)$.
	Therein, the operator $\Lambda_T$ describes how the state at the initial time $t=0$ is transformed into the final state after a time $t=T$.
	Such a process is classical by definition if it maps any classical, \textit{i.e.} incoherent [\textit{cf.} \eref{eq:Incoherent}], input state to a classical output state  \cite{SAWV17,SV15}.
	However, it does not make any statement about any intermediate times, $0<t<T$.
	For instance, it might well be true that $\Lambda_t$ for such an intermediate point in time does indeed generate quantum coherence; see figure \ref{fig:Outline}.

	Therefore, in order to describe a classical evolution which is consistent with incoherent states, we have to derive equations of motions that renders it possible that any classical initial state remains classical for the continuum of all times $0\leq t\leq T$.
	In figure \ref{fig:Outline}, the actual evolution (bottom left) describes a flow of states which does not preserve their classicality.
	Whilst approximations to a time evolution confined to a classical domain are known \cite{H75}, a general method is missing to date.
	Therefore, modified equations of motions are required which correctly map incoherent states onto themselves for all times (bottom right).
	For example, a classical state $|\psi(\boldsymbol q(0))\rangle$ has to propagates as $|\psi(\boldsymbol q(t))\rangle$ for $0\leq t\leq T$.
	In contrast, the typically applied input-output formalism, in which an initially classical state is mapped onto a classical one after a fixed time $T$, does not account for quantum coherence any intermediate time (top row in figure \ref{fig:Outline}), thus not capturing the nonclassical nature of a process as a function of time.
	Beyond previous attempts, we here demand classicality over the entire duration of the process to define a truly classical evolution.

\section{Classical equations of motion in quantum systems}\label{sec:EOM}

	We now derive the sought-after equations of motions using the fundamental principle of least action.
	The goal is to find an evolution that respects the set of classical states for all times.
	For the time being, we focus on pure states; a generalization to mixed states is provided later in section \ref{sec:Mixed}.

	For our parametrization of classical states in terms of generalized, complex coordinates $\boldsymbol q(t)$, we get the time derivative $|\dot\psi\rangle=\dot{\boldsymbol q}\cdot\nabla_{\boldsymbol q}|\psi\rangle=\sum_{j}\dot q_j|\partial_{q_j}\psi\rangle$, which also defines different notations used throughout this work.
	From this, we recast the quantum-physical Lagrangian \eref{eq:Lagrangian} into the form
	\begin{equation}
		\nonumber
		\mathcal L =
		\frac{\rmi \hbar}{2}\sum_{j}\left(
			\dot q_j \langle \psi|\partial_{q_j}\psi\rangle
			-\dot q_j^\ast \langle\partial_{q_j}\psi|\psi\rangle
		\right)
		-\langle\psi|\hat H|\psi\rangle.
	\end{equation}
	The left- and right-hand side of the Euler-Lagrange equation \eref{eq:EulerLagrange} then follow as
	\begin{eqnarray}
		\nonumber
		\frac{\rmd }{\rmd t}\frac{\partial \mathcal L}{\partial \dot q_k^\ast}
		&=&
		\frac{\rmd}{\rmd t}\frac{\hbar}{2\rmi}\langle\partial_{q_k}\psi|\psi\rangle
		\\ &=&
		\frac{\hbar}{2\rmi}\sum_j\left(
			\dot q_j^\ast\langle\partial_{q_j}\partial_{q_k}\psi|\psi\rangle
			{+}\dot q_j\langle\partial_{q_k}\psi|\partial_{q_j}\psi\rangle
		\right),
		\\\nonumber
		\frac{\partial \mathcal L}{\partial q_k^\ast}
		&=&
		\frac{\hbar}{2\rmi}\sum_{j}\left(
			\dot q_j^\ast \langle\partial_{q_k}\partial_{q_j}\psi|\psi\rangle
			-\dot q_j \langle \partial_{q_k}\psi|\partial_{q_j}\psi\rangle
		\right)
		\\ &&
		-\frac{\partial\langle\psi|\hat H|\psi\rangle}{\partial q_k^\ast},
	\end{eqnarray}
	for each component $q_k$.
	This then results in the desired equations of motions which describe the evolution confined to classical states,
	\begin{eqnarray}
		\nonumber
		&& \forall k:
		\frac{\partial\langle\psi|\hat H|\psi\rangle}{\partial q_k^\ast}
		=
		\rmi\hbar\sum_j
		\langle \partial_{q_k}\psi|\partial_{q_j}\psi\rangle
		\dot q_j
		\\ \nonumber
		\Leftrightarrow\quad &&
		\rmi\hbar \langle\nabla_{\boldsymbol q}\psi|\nabla_{\boldsymbol q}\psi\rangle\dot{\boldsymbol q}
		=\nabla_{\boldsymbol q^\ast}\langle\psi|\hat H|\psi\rangle
		\\ \label{eq:ClSE}
		\Leftrightarrow\quad &&
		\langle \nabla_{\boldsymbol q}\psi|\left(
			\rmi\hbar\frac{\rmd}{\rmd t}|\psi\rangle
			-\hat H|\psi\rangle
		\right)
		= \boldsymbol 0,
	\end{eqnarray}
	with $\boldsymbol 0$ denoting the null vector.
	In particular, formula \eref{eq:ClSE} shows that our dynamics follows the tangential space $\mathrm{span}\{|\partial_{q_k}\Psi\rangle:k=0,1,2,\ldots\}$ which is necessary to remain in the set of classical states.
	Thus, we can additionally and equivalently write for \eref{eq:ClSE} that
	\begin{equation}
		\rmi\hbar\frac{\rmd}{\rmd t}|\psi\rangle=\hat H|\psi\rangle+|\chi\rangle,
	\end{equation}
	with $|\chi\rangle\perp\mathrm{span}\{|\partial_{q_k}\psi\rangle:k=0,1,2,\ldots\}$, holds true.
	In this way, our classical evolution resembles a modified Schr\"odinger equation which, however, is restricted to the domain of classical states $|\psi(\boldsymbol q)\rangle$.
	(Note that properties of our equations of motion, such as conservation of energy and normalization, are studied in sections \ref{sec:Properties} and \ref{sec:Mixed}.)

	Specifically, equation \eref{eq:ClSE} forms the basis for our following comprehensive analysis which ultimately enables us to find and characterize quantum properties in the evolution when compared to its classical counterpart.
	Also, these differential equations describe the sought-after classical flow as discussed in figure \ref{fig:Outline}.

\section{Properties and implications}\label{sec:Properties}

	Before applying our newly derived equations of motions to particular examples (section \ref{sec:Applications}), we shall perform an instructive analysis of the implications which follow from our approach.
	This includes certain proof-of-concept examples, functioning as sanity checks, and important implications from a physical perspective.

\subsection{Trivial example}\label{sec:Sanity}

	As a first example, we want to retrieve the actual quantum dynamics under the assumption that all states are classical.
	Whilst not being relevant for quantum coherence, this example serves as a simple test to challenge the capabilities of our method.
	For this purpose, we say $|\psi\rangle=|\psi(\boldsymbol q)\rangle=\sum_{n=1}^{d} q_n|n\rangle$, using an orthonormal basis $\{|n\rangle:n=1,\ldots,d\}$ of a $d$-dimensional Hilbert space.

	We can now apply our previous equations of motions \eref{eq:ClSE}, using $\partial_{q_k}|\psi\rangle=|k\rangle$.
	Then, we readily get the (components of the) Schr\"odinger equation \eref{eq:Schroedinger},
	\begin{equation}
		0=\rmi \hbar \dot q_k-\sum_{n=1}^d \langle k|\hat H|n\rangle q_n,
	\end{equation}
	for all $k=1,\ldots,d$.
	Therefore, this sanity check, which does not exclude any state, correctly yields the expected dynamical behavior.

\subsection{Ehrenfest theorem and Newton's second law}

	As a second, more elaborate example, we study a particle for describing its evolution according to Newton's second law.
	Here, it is convenient to consider cases in which $\boldsymbol q$ describes real-valued parameters, rather than complex ones.
	Following the same treatment leading to \eref{eq:ClSE}, we analogously find
	\begin{equation}
		\label{eq:ClSEreal}
		\boldsymbol 0=\mathrm{Re}\left(\langle \nabla_{\boldsymbol q}\psi|\left[\rmi\hbar\frac{\rmd}{\rmd t}-\hat H\right]|\psi\rangle\right).
	\end{equation}

	Now, we consider transformations of a state which yields a displacement in position and momentum,
	\begin{equation}
		|\psi\rangle=|\psi(\boldsymbol q)\rangle=\hat U(q_x,q_p)|\Phi\rangle,
	\end{equation}
	being defined through a unitary $\hat U=\hat U(q_x,q_p)=\exp(\rmi[q_p\hat x-q_x\hat p]/\hbar)$ and an arbitrary normalized initial state $|\Phi\rangle$.
	Our classical, real parameters are $\boldsymbol q=(q_x,q_p)$ which correspond to a position shift, $\hat U^\dag\hat x\hat U=\hat x+q_x$, and a momentum translation, $\hat U^\dag\hat p\hat U=\hat p+q_p$.
	Furthermore, for this and the following calculations, it is convenient to recall the Baker-Campbell-Hausdorff formula, $\rme^{\hat A+\hat B}=\rme^{-[\hat A,\hat B]/2}\rme^{\hat A}\rme^{\hat B}=\rme^{[\hat A,\hat B]/2}\rme^{\hat B}\rme^{\hat A}$ for operators satisfying the commutation relation $[[\hat A,\hat B],\hat A]=0=[[\hat A,\hat B],\hat B]$.
	
	For applying \eref{eq:ClSEreal}, we first compute the following expressions:
	\begin{eqnarray}
		\partial_{q_x}|\psi(\boldsymbol q)\rangle &=& \frac{1}{\rmi \hbar}\hat U(q_x,q_p)\left(\hat p+\frac{q_p}{2}\right)|\Phi\rangle,
		\\
		\partial_{q_p}|\psi(\boldsymbol q)\rangle &=& -\frac{1}{\rmi \hbar}\hat U(q_x,q_p)\left(\hat x+\frac{q_x}{2}\right)|\Phi\rangle,
		\\
		\frac{\rmd}{\rmd t}|\psi(\boldsymbol q)\rangle &=& \dot q_x\partial_{q_x}|\psi\rangle+\dot q_p\partial_{q_p}|\psi\rangle.
	\end{eqnarray}
	In addition, the Hamilton operator $\hat H=\hat p^2/(2m)+V(\hat x)$, where $V$ describes the potential energy, gives
	\begin{eqnarray}
		\hat H|\psi\rangle
		&=& \hat U\left(
			\frac{(\hat p+q_p)^2}{2m} +V(\hat x+q_x)
		\right)|\Phi\rangle.
	\end{eqnarray}
	After some algebra, and using the above expressions, we can put the classical equations of motion \eref{eq:ClSEreal} for the components $q_x$ and $q_p$ into the form
	\begin{eqnarray}
		\label{eq:Newton1}
		\frac{\rmd}{\rmd t}\left[\langle\hat x\rangle_\Phi+q_x\right] &=& \frac{\langle \hat p\rangle_\Phi+q_p}{m},
		\\\label{eq:Newton2}
		\frac{\rmd}{\rmd t}\left[\langle\hat p\rangle_\Phi+q_p\rangle\right]
		&=&-\frac{\partial \langle V(\hat x+q_x)\rangle_\Phi}{\partial q_x},
	\end{eqnarray}
	respectively, using the notation $\langle\hat A\rangle_\Phi=\langle\Phi|\hat A|\Phi\rangle$, as well as the relations $\hat A\hat B=(\hat A\hat B+\hat B\hat A)/2+[\hat A,\hat B]/2$, $[\hat p,\hat x^m]=-\rmi\hbar m \hat x^{m-1}$, and $[\hat x,\hat p^m]=\rmi\hbar m \hat p^{m-1}$.

	In conclusion, the above expression \eref{eq:Newton2} resembles the classical Newtonian equation of motion, including a force from the movement in a potential $V$ and the definition \eref{eq:Newton1} of the linear momentum.
	The well-known Ehrenfest theorem states the same as we derived here and is, thus, shown to be a special case of our general treatment of the classical evolution in a quantum system.
	Beyond that, we are here additionally able to provide a Schr\"odinger-type equation \eref{eq:ClSEreal} which basically is Newton's second law \eref{eq:Newton2} in disguise [again, including the momentum \eref{eq:Newton1}].
	This establishes a direct connection between classical and quantum dynamics, allowing one to derive classical equations of motions in a full quantum description.

\subsection{Energy conservation}

	Beyond specific examples, we now prove an essential feature of our dynamics, stating that our classical evolution obeys the energy conservation.
	Thus, we compute the time derivative of the expectation value of the Hamilton operator for classical states, $\langle\psi|\hat H|\psi\rangle$.
	Using $\langle\partial_{q_k}\psi|\hat H|\psi\rangle=\rmi\hbar\langle\partial_{q_k}\psi|\dot\psi\rangle$ [\textit{cf.} \eref{eq:ClSE}], we find
	\begin{eqnarray}
		\nonumber
		\frac{\rmd\langle\psi|\hat H|\psi\rangle}{\rmd t}
		&=& \sum_k\left(
			\dot q_k^\ast\langle\partial_{q_k}\psi|\hat H|\psi\rangle
			+\dot q_k\langle\psi|\hat H|\partial_{q_k}\psi\rangle
		\right)
		\\\nonumber
		&=&
		\rmi \hbar \sum_{k}\dot q_k^\ast \langle\partial_{q_k}\psi|\dot\psi\rangle
		-\rmi \hbar \sum_{k}\dot q_k \langle\dot\psi|\partial_{q_k}\psi\rangle
		\\
		&=& \rmi \hbar \langle\dot\psi|\dot\psi\rangle-\rmi \hbar \langle\dot\psi|\dot\psi\rangle
		=0.
	\end{eqnarray}
	This simple corollary to our approach demonstrates that our classical dynamics is in fact compatible with the fundamental conservation law of energy.
	Again, the same can be shown for real parameters $\boldsymbol q$ in the same manner.

\section{Evolution of mixed states in closed systems}\label{sec:Mixed}

\subsection{Mixed states and pure auxiliary states}

	In our previous derivation, we focused on pure states.
	However, the set of classical states describes a convex combination of pure states which includes mixed ones, \textit{cf.} \eref{eq:Incoherent}.
	In general, it is convenient to represent mixed states $\hat\rho$ via pure ones by attaching a second bath-type, auxiliary system, $|\Psi\rangle=\sum_{n}|\psi_n\rangle\otimes|n\rangle_\mathrm{aux}$, where $|n\rangle_\mathrm{aux}$ form an orthonormal basis.
	When tracing over the second system, we get any desired mixed state,
	\begin{equation}
		\hat\rho=\mathrm{tr}_\mathrm{aux}\left(|\Psi\rangle\langle\Psi|\right)=\sum_n |\psi_n\rangle\langle\psi_n|,
	\end{equation}
	where the normalization $p_n=\langle\psi_n|\psi_n\rangle$ defines the probability such that $\mathrm{tr}(\hat\rho)=1$.
	Note that, if required, the discrete sum can be replaced with an integral.

	For simplicity, we restrict our consideration in this work to closed systems, implying no interaction between the bath and the actual state.
	In this scenario, the Hamilton operator reads as $\hat H=\hat H_0\otimes\hat 1_\mathrm{aux}$.
	Thus, the Lagrangian \eref{eq:Lagrangian} decomposes as
	\begin{equation}
		\mathcal L=\sum_n\left[
			\frac{\rmi\hbar}{2}\left(
				\langle\psi_n|\dot\psi_n\rangle{-}\langle\dot\psi_n|\psi_n\rangle
			\right)
			-\langle\psi_n|\hat H_0|\psi_n\rangle
		\right].
	\end{equation}
	Because of this structure, the action $\mathcal S$ also becomes a sum of the individual components for $|\psi_n\rangle$.
	Thus, a separate optimization over each component is possible, and the results of the previous section can be applied for each $|\psi_n(\boldsymbol q)\rangle$ separately.
	Consequently, the classical evolution of the statistical mixture $\hat \rho$ is simply given by
	\begin{equation}
		\hat\rho(t)=\sum_{n}|\psi_n(\boldsymbol q(t))\rangle\langle \psi_n(\boldsymbol q(t))|.
	\end{equation}

\subsection{Normalization and probabilities}\label{subsec:Phase}

	For each component of the statistical ensemble $\hat\rho$, the normalization corresponds to the probability $p_n=\langle\psi_n|\psi_n\rangle$, as defined earlier.
	Thus, it is reasonable to investigate the impact of a scale factor for the state.
	Note that, in the following, we drop the index $n$.

	We now consider the classical state in the form
	\begin{equation}
		|\psi\rangle=q_0|\Psi(q_1,q_2,\ldots)\rangle=q_0|\psi'\rangle,
	\end{equation}
	where $q_0$ defines a preceding factor which accounts for normalization and a global phase.
	In addition, the remaining vector reads $|\psi'\rangle=|\psi'(\boldsymbol q')\rangle$ for the parameter tuple $\boldsymbol q'=(q_1,q_2,\ldots)$, excluding the zeroth entry.

	Taking this decomposition above into account, the Lagrangian \eref{eq:Lagrangian} rewrites as
	\begin{equation}
		\mathcal L=\frac{\rmi \hbar}{2}\left(q_0^\ast\dot q_0-\dot q_0 q_0^\ast\right)\langle\psi'|\psi'\rangle+|q_0|^2\langle\psi'|\psi'\rangle\mathcal L',
	\end{equation}
	in which we define the real-valued function
	\begin{equation}
		\label{eq:LagrangianNormalized}
		\mathcal L'
		=\frac{\rmi \hbar(\langle\psi'|\dot\psi'\rangle-\langle\dot\psi'|\psi'\rangle)}{2\langle\psi'|\psi'\rangle}
		-\frac{\langle\psi'|\hat H|\psi'\rangle}{\langle\psi'|\psi'\rangle},
	\end{equation}
	describing another Lagrangian.
	The Euler-Lagrange equation \eref{eq:EulerLagrange} for $\mathcal L$ and $q_0$ can be put into the form
	\begin{equation}
		0=\frac{\dot q_0}{q_0}+\frac{1}{2}\frac{\rmd \ln\langle\psi'|\psi'\rangle}{\rmd t}+\frac{\mathcal L'}{\rmi \hbar},
	\end{equation}
	which can be solved analytically,
	\begin{equation}
		q_0(t)=C\frac{
			\exp\left[
				-\frac{1}{\rmi \hbar}\int_0^t \rmd t'\mathcal L' 
			\right]
		}{\sqrt{\langle\psi'|\psi'\rangle}},
	\end{equation}
	with a constant $C$.
	Importantly, one can conclude from this solution that the conservation normalization applies because
	\begin{equation}
		\langle\psi|\psi\rangle=|q_0|^2\langle\psi'|\psi'\rangle=|C|^2
	\end{equation}
	holds true.
	In fact, this proves the time-independence of the probability $p=\langle\psi|\psi\rangle$, being identical to the behavior of closed quantum systems without the restriction to a classical subset of states.

	For the remaining parameters $\boldsymbol q'$, we may now evaluate our general equation of motion \eref{eq:ClSE},
	\begin{eqnarray}
		\nonumber
		0&=&\rmi \hbar q_0^\ast\dot q_0\langle \partial_{q_k}\psi'|\psi'\rangle
		\\&&
		+|q_0|^2\langle\partial_{q_k}\psi'|\left[
			\rmi\hbar\frac{\rmd}{\rmd t}|\psi'\rangle-\hat H|\psi'\rangle
		\right]
	\end{eqnarray}
	for $k\neq0$.
	Applying the solution for $\dot q_0$, we then get
	\begin{equation}
		\label{eq:EOMnonNorm}
		\boldsymbol 0=\langle \nabla_{\boldsymbol q'}\psi'|
		\left[\hat 1-\frac{|\psi'\rangle\langle\psi'|}{\langle\psi'|\psi'\rangle}\right]
		\left[
			\rmi \hbar\frac{\rmd}{\rmd t}-\hat H
		\right]|\psi'\rangle.
	\end{equation}
	It is worth mentioning that those equations of motion are the same as obtained from our equations \eref{eq:ClSE} when replacing the Lagrangian with $\mathcal L'$ [\textit{cf.} \eref{eq:LagrangianNormalized}], which, in turn, means automatically an restriction to normalized states, $|\psi\rangle=\langle\psi'|\psi'\rangle^{-1/2}|\psi'\rangle$, since we have $\mathrm{Im}(\langle\psi|\dot\psi\rangle)=\mathrm{Im}(\langle\psi'|\dot\psi'\rangle/\langle\psi'|\psi'\rangle)$ in this case.

\section{Applications}\label{sec:Applications}

	To this end, we derived equations of motions to describe a classical flow in evolving quantum systems, we showed that the Schr\"odinger equation and Newton's equations of motion follow from our general treatment, and we analyzed other features and extensions of this new method as well.
	In this section, we focus on studying several examples to demonstrate the broad applicability of our approach to various physical systems.
	In particular, we consider nonlinear processes in quantum optics (section \ref{sec:optics}), the dynamical description of semiclassical models (section \ref{sec:semiclassical}), and interacting multipartite systems for characterizing the entanglement dynamics (section \ref{sec:entanglement}).

	The choice of a set of classical reference states depends on the physical scenario under study.
	A rather general approach to construct such classical states is formulated via Lee groups \cite{P72}.
	This, for example, leads to the well-known classical states of the harmonic oscillator \cite{S26,H85}, being displaced vacuum states.
	Generalizations to nonlinear oscillators systems have been constructed in a similar fashion \cite{MV96,MMSZ97}.
	For implementations of a variety of quantum information processing protocols, different kinds of quantum coherence are required \cite{SAP17}, thus leading to various notions of classically correlated states.
	Amongst the most prominent forms of quantum coherence is its nonlocal manifestation, resulting in the notion of entanglement \cite{W89}.
	Consequently, a manifold of methods have been developed to certify this quantum effect for single points in time (\textit{e.g.} see \cite{HHHH09,GT09} for reviews).
	Recently, the involved interplay of second quantization and entanglement has been studied \cite{M17,SPBS19}, potentially leading to further insights into the quantum-classical transition \cite{M15}.

	In the following, we study quantum features in the time evolution which apply to different concepts of quantumness.

\subsection{Nonlinear quantum optics}\label{sec:optics}

	A coherent state $|\alpha\rangle=\rme^{-|\alpha|^2/2}\sum_{n=0}^\infty \alpha^n |n\rangle/\sqrt{n!}$ (where we can set $\boldsymbol q=\alpha$) behaves most similarly to a classical harmonic oscillator \cite{S26}.
	A monochromatic radiation mode, represented by the annihilation operator $\hat a$, is an example of a quantum-physical harmonic oscillator in which the coherent amplitude $\alpha$ then relates to the classical electromagnetic field \cite{TG65,M86}.
	(Namely, we have $\alpha=(\varepsilon_0^{1/2} E+\rmi\mu_0^{-1/2}B)/\sqrt{2\hbar\omega}$, where $E$ and $B$ are the electric and magnetic fields and $\varepsilon_0$ and $\mu_0$ are the permittivity and permeability of vacuum.)
	In order to avoid confusions between historical notations and modern concepts, please note that the word ``coherent'' in coherent state---being the classical reference---relates to classical coherence in optics rather than quantum coherence.

	Let us focus on a single mode as the generalization to multimode light is straightforward.
	A general nonlinear Hamilton operator expands in its normally ordered form (\textit{i.e.} in the basis $\hat a^{\dag k}\hat a^{l}$, where $\hat a^0=\hat 1$) as $\hat H=\hbar\sum_{k,l=0}^\infty \Omega_{k,l}\hat a^{\dag k}\hat a^{l}$ \cite{VW06}.
	From $\hat a^l|\alpha\rangle=\alpha^l|\alpha\rangle$, we now also get a classical Hamiltonian in Taylor expansion,
	\begin{equation}
		\mathcal H=\langle\alpha|\hat H|\alpha\rangle=\hbar\sum_{k,l=0}^\infty \Omega_{k,l}\alpha^{\ast k}\alpha^{l}.
	\end{equation}
	Using this expression, we can evaluate our classical equation of motion \eref{eq:ClSE} for the parameter $\alpha$.
	This yields
	\begin{eqnarray}
		\label{eq:GenCohrent}
		\rmi\frac{\rmd}{\rmd t}\alpha=\sum_{k,l=0}^\infty \Omega_{k,l}\,k\,\alpha^{\ast (k-1)}\alpha^{l}=\frac{1}{\hbar}\frac{\partial\mathcal H}{\partial \alpha^\ast},
	\end{eqnarray}
	where the right-most expression is a convenient rewriting of the prior term.
	(See \ref{app:Hybrid} for further details.)
	This classical equation can be compared to the Heisenberg equation for the quantized radiation field,
	\begin{eqnarray}
		\label{eq:GenAnnihilation}
		\rmi\frac{\rmd}{\rmd t}\hat a=\frac{1}{\hbar}[\hat a,\hat H]=\sum_{k,l=0}^\infty \Omega_{k,l}\,k\,\hat a^{\dag(k-1)}\hat a^l.
	\end{eqnarray}
	Therein, we applied $[\hat a,\hat a^{\dag k}]=k\hat a^{\dag(k-1)}$, which enables us to formally identify $[\hat a,\hat H]=\partial\hat H/\partial \hat a^\dag$.
	This means that our classical nonlinear wave equation \eref{eq:GenCohrent} takes formally the same form as the Heisenberg equation \eref{eq:GenAnnihilation}.
	However, the classical dynamics is quite different to the quantum evolution because of the noncommuting nature of quantum-optical operators.

	To demonstrate this, let us study the propagation of light in a Kerr medium, given by the Hamilton operator \cite{YS86,M86a}
	\begin{eqnarray}
		\hat H=\hbar\omega\hat a^\dag\hat a+\frac{\hbar\kappa}{2}\hat a^{\dag 2}\hat a^2,
	\end{eqnarray}
	where $\kappa$ is the coupling constant.
	Consequently, the classical Hamiltonian reads
	\begin{eqnarray}
		\mathcal H=\hbar\omega|\alpha|^2+\frac{\hbar\kappa}{2}|\alpha|^4.
	\end{eqnarray}
	Applying our method and the Heisenberg equation, we get the following equations of motion for the classical and quantum domain:
	\begin{eqnarray}
		\rmi\frac{\rmd}{\rmd t}\alpha&=&\omega(1+\chi\alpha^\ast\alpha)\alpha,
		\\
		\rmi\frac{\rmd}{\rmd t}{\hat a}&=&\omega(\hat 1+\chi\hat a^\dag\hat a)\hat a,
	\end{eqnarray}
	respectively, where $\chi=\kappa/\omega$ quantifies the intensity dependent contribution to the refractive index.
	Moreover, the solution of the classical dynamics reads
	\begin{eqnarray}
		\label{eq:KerrClassSolve}
		\alpha(t)=\rme^{-\rmi t[\omega +\kappa |\alpha(0)|^2]}\alpha(0).
	\end{eqnarray}
	Similarly, the quantum field propagates according to
	\begin{eqnarray}
		\label{eq:KerrQuantSolve}
		\hat a(t)=\rme^{-\rmi t[\omega \hat 1+\kappa \hat n(0)]}\hat a(0),
	\end{eqnarray}
	where $\hat n=\hat a^\dag\hat a$ is the photon-number operator.

	The classical position in phase space is given by equation \eref{eq:KerrClassSolve}.
	By contrast, for the initial state $|\alpha(0)\rangle$, the quantum solution \eref{eq:KerrQuantSolve} yields
	\begin{eqnarray}
		\label{eq:KerrQuantSolveMean}
		\langle\alpha(0)|\hat a(t)|\alpha(0)\rangle=\rme^{-\rmi\omega t+(\rme^{-\rmi\kappa t}-1)|\alpha(0)|^2}\alpha(0),
	\end{eqnarray}
	using $\rme^{x\hat n}|\alpha\rangle=|\rme^{x}\alpha\rangle$ and $\langle\alpha|\alpha'\rangle=\rme^{-(|\alpha|^2+|\alpha'|^2)/2}\rme^{\alpha^\ast\alpha'}$.
	For small times, implying $\rme^{-\rmi\kappa t}-1\approx -\rmi\kappa t$, the classical position in phase space coincides with the quantum one.
	However, this is not true for all time, allowing us to discern the quantum and classical propagation.
	Also, we can apply the classical solution even to an initially nonclassical state.
	(Specifically, for a classical evolution of a superposition state, the solutions of the classical dynamics, $\alpha_k(t)$, apply to a superposition of coherent states as $|\psi(t)\rangle=\sum_{k}\psi_k|\alpha_k(t)\rangle$.)
	This further enables us to distinguish quantum effects resulting from a nonclassical dynamics from nonclassicality which was already present at the initial time, $t=0$.

	The solution \eref{eq:KerrClassSolve} describes the evolution of the system constrained to classical, \textit{i.e.} coherent, states.
	By contrast, the nonidentical result in Eq. \eref{eq:KerrQuantSolveMean} can be obtained from solving the quantum-physical Heisenberg equation \eref{eq:GenAnnihilation} for the considered Kerr Hamilton operator and computing the time-dependent expectation value for coherent states.
	A typical \textit{ad hoc} approach to estimate classical processes is the factorization of quantum-physical expectations values, $\langle \hat a^{\dag k}\hat a^{l}\rangle=\langle\hat a^\dag\rangle^{k}\langle\hat a\rangle^l$, which yields the same classical solution $\alpha(t)$ from the Heisenberg equation.
	Here, however, our method renders it possible to avoid such an assumption and base the classical evolution on equations of motions derived from first principles.

\begin{figure}
	\includegraphics[width=\columnwidth]{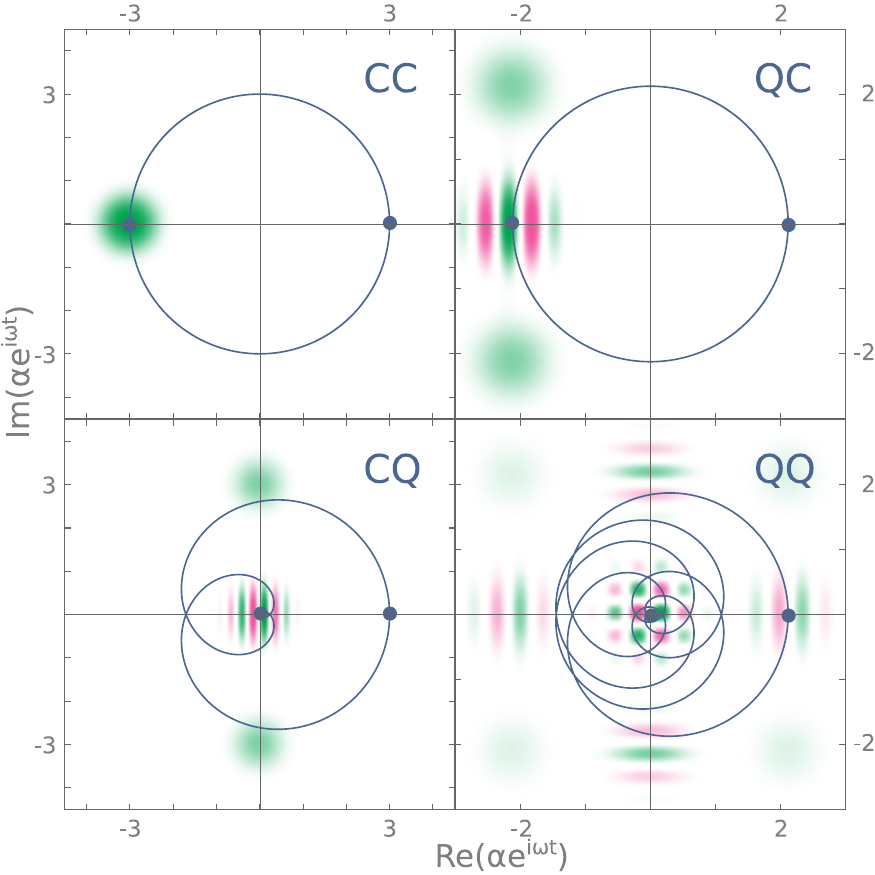}
	\caption{
		Phase-space representation at $t=\pi/\kappa$, rotated by $e^{i\omega t}$ to counter the free evolution.
		The density plot (green and magenta parts) depict the interfering (positive and negative) contributions in the Wigner function of the evolved state.
		The first and second letter of a panel's label indicates the type of initial state and dynamics, respectively; both can be either classical (C) or quantum (Q).
		The curves depict the trajectory of the mean value, $\langle \hat a\rangle \rme^{\rmi\omega t}$.
		Bullet points mark the initial time $t=0$ (right) and an intermediate time $t=\pi/(2\kappa)$ (left).
	}\label{fig:Kerr}
\end{figure}

	In figure \ref{fig:Kerr}, we compare the classical (top row) and quantum (bottom row) dynamics for states which can be either classical (left column) or nonclassical (right column) for $t=0$.
	The trajectories (solid curves) significantly differ from one another, \textit{cf.} \eref{eq:KerrClassSolve} and \eref{eq:KerrQuantSolveMean}.
	Thus, the quantum trajectory exhibits characteristics (here, additional loops) not present in its classical counterpart.

	In addition, the Wigner function of the evolved states for $t\kappa=\pi/2$ is depicted in figure \ref{fig:Kerr}.
	Negativities of the Wigner function (magenta areas) are a sufficient indication of the nonclassicality in terms of the Glauber-Sudarshan function \cite{G63,S63}, which is the actual figure of merit but is itself highly singular in the present scenario \cite{SKVGZB13}; see also \cite{MRV98,AKM12} for a different analysis of the evolution in phase space.
	The classical evolution preserves---up to a $\pi$ rotation---the features of the initial state, \textit{cf.} panels CC for $|\alpha(0)=3\rangle$ and QC for the superposition $|\psi(0)\rangle\propto |3\rme^{-\rmi\pi/4}\rangle-|3\rme^{\rmi\pi/4}\rangle$.
	In case of the quantum evolution, nonclassicality is generated via the dynamics itself, even for the initially classical state (panel CQ) and additional nonclassical interference patterns occur for the initially nonclassical state (panel QQ).
	Thus, we can clearly discriminate quantum effects originating from the dynamics from those originating from the initial nonclassicality.

\subsection{Semiclassical models}\label{sec:semiclassical}

	Semiclassical models are another application of our technique.
	Such hybrid systems are important for coupling a classical degree of freedom to a quantum-mechanical one, \textit{e.g.} to describe measurements with macroscopic devices.
	However, the quantum evolution typically yields unphysical results as quantum properties can ``leak'' into the classical domain because of the interaction;
	see \cite{LG18} for a recent study and \cite{E12,FLE14} for quantum-classical hybrid systems.
	Our approach can overcome such deficiencies.

	As a proof of concept, we study a light field interacting with a two-level atom using the Jaynes-Cummings model \cite{JC63},
	\begin{eqnarray}
		\nonumber
		\frac{\hat H}{\hbar}
		=&\omega\hat a^\dag \hat a\otimes\hat 1+\omega \hat 1\otimes|e\rangle\langle e|
		\\\label{eq:JCHamiltonOp}
		& +\rmi\kappa\hat a\otimes|e\rangle\langle g|-\rmi\kappa\hat a^\dag\otimes|g\rangle\langle e|,
	\end{eqnarray}
	were $|e\rangle$ and $|g\rangle$ are the excited and ground state of the atom and $\kappa$ is a positive coupling constant.
	Here, the resonant scenario is considered in which the energy difference between the atomic states is proportional to the frequency of the field, $\hbar\omega$.

	To treat the system in a semiclassical framework, we say that a classical coherent state $|\alpha\rangle$ represents a classical electromagnetic field mode.
	The state of the atom, $|\phi\rangle=\phi_g|g\rangle+\phi_e|e\rangle$, is treated in the quantum description.
	Thus, we have to restrict to states $|\alpha\rangle\otimes|\phi\rangle$ together with a parametrization $\boldsymbol q=(\alpha,|\phi\rangle)$.

	Applying the equation of motion \eref{eq:ClSE} to this Hamiltonian for each component of the pair $\boldsymbol q=(\alpha,|\phi\rangle)$, we find
	\begin{eqnarray}
		\label{eq:JCclassical}
		\rmi\frac{\rmd}{\rmd t}\alpha
		&=&
		\omega\alpha-\rmi\kappa\langle\phi|g\rangle\langle e|\phi\rangle,
		\\\label{eq:JCquantum}
		\rmi\hbar\frac{\rmd}{\rmd t}|\phi\rangle
		&=&\Big[
			\hbar\omega|e\rangle\langle e|
			{+}\rmi\hbar\kappa\alpha|e\rangle\langle g|
			{-}\rmi\hbar\kappa\alpha^\ast|g\rangle\langle e|
		\Big]|\phi\rangle.
	\end{eqnarray}
	A detailed derivation of these equations is provided in \ref{app:Hybrid}, where a more general hybrid system (including a $d$-level system and arbitrary interactions) is presented.
	On the one hand, equation \eref{eq:JCclassical} is a classical wave equation, which includes a source term proportional to the scalar, time-dependent function $\langle\phi(t)|g\rangle\langle e|\phi(t)\rangle$.
	On the other hand, the dynamics of the quantum atom is described in terms of the Schr\"odinger equation \eref{eq:JCquantum}, including an effective interaction proportional to the operator $\rmi\alpha(t)|e\rangle\langle g|-\rmi\alpha(t)^\ast|g\rangle\langle e|$.
	We emphasize that the classical and quantum components continuously interact with each other.
	Still, because our equations of motion restrict the flow to states of the form $|\alpha\rangle\otimes|\phi\rangle$, the solutions $\alpha(t)$ of the classical wave equation \eref{eq:JCclassical} remain classical field amplitudes at all times $t$, in contrast to earlier approaches.

	As an example, we say that the atom is initially in the quantum superposition state $|\phi(0)\rangle=(|g\rangle+|e\rangle)/\sqrt2$ and the light field in the vacuum state, $\alpha(0)=0$.
	The actual Schr\"odinger equation \eref{eq:Schroedinger} for the fully quantum Hamilton operator \eref{eq:JCHamiltonOp} is solved by
	\begin{eqnarray}
		\nonumber
		|\psi(t)\rangle
		&=&
		\frac{
			|n{=}0\rangle
			{-}\rme^{-\rmi\omega t}\sin(\kappa t)|n{=}1\rangle
		}{\sqrt2}{\otimes}|g\rangle
		\\\label{eq:JCsolutionquantum}
		&& +\frac{\rme^{-\rmi\omega t}\cos(\kappa t)|n{=}0\rangle}{\sqrt2}{\otimes}|e\rangle,
	\end{eqnarray}
	where $|n\rangle$ denotes the $n$th photon-number state, which is nonclassical for $n>0$ \cite{VW06}.
	Note that, for $\sin(\kappa t)\neq0$, this solution describes an entangled state.
	More generally, the joint quantum system evolves into a state in which the optical mode cannot be considered as classical at every time, despite being classical for certain points in time which satisfy $\kappa t=0\,\mathrm{mod}\,\pi$.

	By contrast, our semiclassical equations of motion, \eref{eq:JCclassical} and \eref{eq:JCquantum}, yield a semiclassical state for all times, $|\alpha(t)\rangle\otimes|\phi(t)\rangle$, where the classical field amplitude reads
	\begin{eqnarray}
		\alpha(t)=-\frac{\rme^{-\rmi\omega t}\sin[\vartheta\big(\kappa t/\sqrt2\big)]}{\sqrt2},
	\end{eqnarray}
	and the factorized quantum state of the atom reads
	\begin{eqnarray}
		\nonumber
		|\phi(t)\rangle
		&=&\frac{
			\sqrt{1{+}\sin^2[\vartheta\big(\kappa t{/}\!\sqrt2\big)]}|g\rangle
		}{\sqrt2}
		\\
		&&+\frac{
			\rme^{-\rmi\omega t}\cos[\vartheta\big(\kappa t{/}\!\sqrt2\big)]|e\rangle
		}{\sqrt2}.
	\end{eqnarray}
	Therein, $\vartheta(x)$ denotes a Jacobi elliptic function defined by $\rmd\vartheta/\rmd x=(1+\sin^2\vartheta)^{1/2}$ and $\vartheta(0)=0$, which can be approximated with $\vartheta(x)\approx\mu x$ ($\mu=1.198\,140\,\ldots$).
	In these solutions, the classically evolved state keeps its tensor-product form $|\alpha(t)\rangle\otimes|\phi(t)\rangle$ for all times $t$ in which the first component describes a classical wave, in contrast to the full quantum solution \eref{eq:JCsolutionquantum} which is unable to model a semiclassical system.

	Thus, our method renders it possible to formulate self-consistent semiclassical models for studying the joint evolution of a classical and quantum system, including interactions between them and for all times.
	Assumptions, approximations, and restrictions---such as applied to derive the Maxwell-Bloch equations for the considered scenario---become superfluous.
	In addition, inconsistencies which lead to unphysical solutions, \textit{e.g.} as previously discussed in \cite{LG18}, do not occur in our approach.

\subsection{Multipartite entanglement}\label{sec:entanglement}

	After considering a bipartite system and as a final application, further demonstrating the broad usefulness of our method, we can use our approach to study multipartite quantum correlations.
	An $N$-partite quantum system can be decomposed into $K$ parties, each including $N_j$ individual subsystems (with $j=1,\ldots,K$ and $N_1+\cdots+N_K=N$).
	For instance, the composite four-party system $\{A,B,C,D\}$ can be decomposed into a bipartition of the form $\{A,B\}{:}\{C,D\}$, \textit{i.e.} $N=4$, $K=2$, and $N_1=N_2=2$.
	With an increasing number of parties, the number of possible decompositions (thus, potential quantum correlations) increases exponentially, which already makes a time-independent characterization of multipartite correlations a sophisticated problem \cite{HHHH09}.
	
	Let us again begin with identifying the classical reference in terms of pure states.
	A pure state is $K$-separable with respect to a given partition if it can be written in a tensor product form, $|\psi_1\rangle\otimes\cdots\otimes|\psi_K\rangle$.
	A mixed separable state is a statistical mixture of pure separable states \cite{W89}, which includes classical correlations only.
	If a state cannot be described in this form, it is referred to as $K$-entangled.
	Consequently, for the purpose of this application, we identify the notion of classical states with $K$-separable ones.

	In \cite{SW17a}, we derived and characterized a form of evolution which is restricted to separable states.
	In fact, applying the technique formulated here---equation \eref{eq:ClSE} with $\boldsymbol q=(|\psi_1\rangle,\ldots,|\psi_K\rangle)$---we find the same equation of motions for $K$-separable states, which can be put into the form
	\begin{eqnarray}
		\label{eq:SSE}
		\rmi\hbar\frac{\rmd}{\rmd t}|\psi_j\rangle=\hat H_{\psi_1,\ldots,\psi_{j-1},\psi_{j+1},\ldots,\psi_K}|\psi_j\rangle,
	\end{eqnarray}
	for $j=1,\ldots,K$ and using so-called partially reduced Hamilton operators,
	\begin{eqnarray}
		&\hat H_{\psi_1,\ldots,\psi_{j-1},\psi_{j+1},\ldots,\psi_K}
		\\\nonumber
		=&\big(\langle \psi_{1}|\otimes\cdots\otimes\langle\psi_{j-1}|\otimes\hat 1_j\otimes\langle \psi_{j+1}|\otimes\cdots\otimes\langle\psi_{K}|\big)\hat H
		\\\nonumber
		&\times\big(|\psi_{1}\rangle\otimes\cdots|\psi_{j-1}\rangle\otimes\hat 1_j\otimes|\psi_{j+1}\rangle\otimes\cdots|\psi_{K}\rangle\big),
	\end{eqnarray}
	where $\hat 1_j$ denotes the identity of the $j$th subsystem.
	We refer to equation \eref{eq:SSE} as separability Schr\"odinger equation.
	The consistency with earlier results in \cite{SW17a} demonstrates that the method developed here encompasses yet another technique as a special case.

	Going far beyond previously studied bipartite examples, including those in \cite{SW17a}, we now consider the multipartite entanglement dynamics of a system which consists of $N$ coupled, quantum-mechanical oscillators, defined through the Hamilton operator
	\begin{eqnarray}
		\label{eq:HmechOsci}
		\hat H=
		\sum_{k=1}^N\left(\frac{1}{2m}\hat p_k^2+\frac{m\omega^2}{2}\hat x_k^2\right)
		+\frac{\kappa}{2}\sum_{k>l}(\hat x_k-\hat x_l)^2.
	\end{eqnarray}
	Here, $\hat x_k$ and $\hat p_k$ represent the position and momentum of the $k$th oscillator, respectively, with a mass $m$ and an eigenfrequency $\omega$.
	The second sum in the Hamilton operator includes the interaction which couples all individual oscillators to each other.

	Similarly to the separability Schr\"odinger equation \eref{eq:SSE} for the state, we can also establish Heisenberg-type equations of motion.
	In our case, we get
	\begin{eqnarray}
		\rmi\hbar\frac{\rmd\vec{\hat x}_j}{\rmd t}
		&=&[\vec{\hat x}_j,\hat H_{\psi_1,\ldots,\psi_{j-1},\psi_{j+1},\ldots,\psi_K}],
		\\
		\rmi\hbar\frac{\rmd\vec{\hat p}_j}{\rmd t}
		&=&[\vec{\hat p}_j,\hat H_{\psi_1,\ldots,\psi_{j-1},\psi_{j+1},\ldots,\psi_K}],
	\end{eqnarray}
	for $j=1,\ldots,K$ and $\vec{\hat x}_j$ ($\vec{\hat p}_j$) containing all positions (momenta) of the $j$th subsystem.
	We analytically solve these equations for the Hamilton operator \eref{eq:HmechOsci} to describe the dynamics of the covariance matrix for arbitrary $K$ partitions and any initial state.
	In contrast to our previous calculations, solving those equations is more sophisticated;
	thus, we provide a detailed calculation in \ref{app:Entanglement}.

	As a concrete example, we study the case in which the initial state is the thermal state of $\hat H$ of the Hamilton operator \eref{eq:HmechOsci} without interactions, $\kappa=0$.
	This yields for $t=0$ a mixed, continuous-variable, $N$-fold tensor-product state, which is separable.
	For the propagation in time, we then introduce an interaction between all particles, $\kappa\neq0$.
	Furthermore, we particularly want to follow the evolution of the variance of the mean momentum operator,
	\begin{eqnarray}
		\hat P=\frac{\hat p_1+\cdots+\hat p_N}{N},
	\end{eqnarray}
	to which we apply our exact solutions from \ref{app:Entanglement}.
	That is, for $t\geq 0$ and a nonvanishing coupling, $\kappa>0$, we obtain the $K$-separable evolution of the quantity under study as
	\begin{eqnarray}
		\nonumber
		V_{P}(t)
		= \langle(\Delta\hat P)^2\rangle
		&=& \frac{\hbar m\omega}{N}\left[\frac{1}{2}+\frac{\rme^{-\hbar\omega/(k_BT)}}{1-\rme^{-\hbar\omega/(k_BT)}}\right]
		\\\label{eq:PVariance}
		&&{\times}
		\left[1+r_K\sin^2\left(\sqrt{1{+}r_K}\omega t\right)\right],
	\end{eqnarray}
	for a balanced partitioning (\textit{i.e.} $N_1=\cdots=N_K=N/K$), the temperature $T$, the Boltzmann constant $k_B$, and the parameter $r_K=(1-K^{-1})N\kappa/(m\omega^2)$.

\begin{figure}
	\includegraphics[width=\columnwidth]{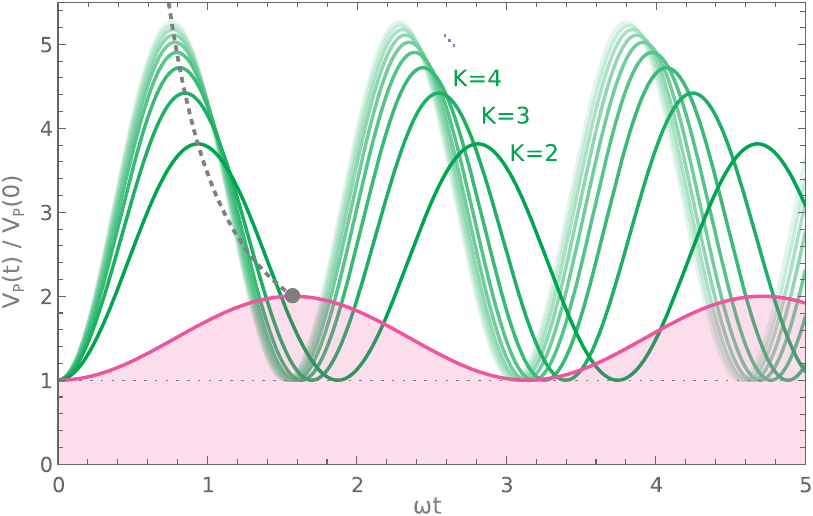}
	\caption{
		Comparison of $K$-separable evolutions (increasingly lighter green curves, labeled with $K=2,\ldots, 10$) with the entanglement-generating dynamics (bottom, magenta curve for $K=1$) for the relative variance of $\hat P$ [\textit{cf.} \eref{eq:PVariance}].
		The system consists of $N=10!$ harmonic oscillators with a relative coupling constant $\kappa/(m\omega^2)=10^{-6}$.
		The dashed (gray) curve indicates maximal amplitudes which are attained for one half period of the $K$-separable evolution.
	}\label{fig:MultiEnt}
\end{figure}

	In figure \ref{fig:MultiEnt}, we show the evolution of the relative variance $V_{P}(t)/V_{P}(0)$ for different $K$ partitions of a macroscopic ensemble consisting of $N=10!$, \textit{i.e.} about 3.6 million, individual oscillators;
	see \cite{SV13,SW17} for a related but time-independent analysis.
	Because of the rescaling with the initial variance, the depicted curves apply to all initial temperatures which define the thermal states at $t=0$.
	For all times, the variances never fall below the initial value, $V_{P}(t)/V_{P}(0)\geq 1$, which eliminates the possibility of squeezing as a single-time quantum effect.
	Even in the weak coupling regime chosen for figure \ref{fig:MultiEnt}, we can clearly see distinct differences in the evolution via the Schr\"odinger equation (magenta curve) and the $K$-separable dynamics (green curves) which are confined to classically correlated states.

	In particular, the actual entanglement-generating evolution (magenta curve), also resembling the trivial partition with $K=1$, has the smallest initial increment of the variance.
	This is clearly different when compared to the bipartition, $K=2$, separating the entire ensemble of oscillators into two parts with $N_1=N_2 \approx 1.8\times10^6$ particles in each subsystem.
	In fact, with increasing separation $K$ (gathering fewer particles in each subsystem, $N_1=\cdots=N_K=N/K$), the speed with which $V_{P}(t)/V_{P}(0)$ increases at $t=0$ increases as well.
	Also, the difference between the $K$-separable dynamics and the actual propagation becomes increasingly pronounced in the amplitude of the oscillation, as indicated by the dashed gray curve.
	Moreover, the time scales (\textit{i.e.} periods) also allow us to discern the processes for different degrees $K$ of separation.
	Therefore, our general method enables us to study distinct features in the dynamics of entangled and the classically correlated systems (for arbitrary $K$) which further enable us to identify measurable signatures of a quantum correlated process.

\section{Summary, discussion, and conclusion}\label{sec:Conclusion}

	In summary, we devised a methodology to describe classical dynamics in quantum systems.
	This was achieve by deriving equations of motions from first principles which confine the evolution of a system to a classical domain.
	This approach further allowed us to compare the quantum dynamics with its classical counterpart for the purpose of identifying quantum properties in the system's propagation in time.
	In contrast to simple input-output relations, our technique is not limited to the initial and final point in time and, in fact, renders it possible to monitor the quantum or classical character of a process over the continuum of its duration.
	Moreover, our framework applies to a rather general family of classically motivated concepts, as we demonstrated with several examples.

	Based on the fundamental principle of least action, which applies to any physical systems and dynamics which include general constraints, we derived our equations of motions for the parameters which characterize a given set of classical states.
	Consequently, the solutions of our equations of motion are confined to the manifold of classical states for all times.
	Whenever the actual quantum dynamics differs from the classical behavior, the quantum nature of the process is verified.
	We also showed how our approach naturally extends to mixed classical states and how it enables us to derive Newton's second law from a quantum-physical Schr\"odinger-type equation.

	To even further underline the power of our framework, we considered several examples with relevance in physics.

	In our first example, we analyzed processes in nonlinear quantum optics.
	We demonstrated that our classical equations of motion and the quantum-physical Heisenberg equations formally share the same structure.
	However, their solutions exhibit distinguishing features, which are not accessible with other methods which approximate a classical dynamics, such as mean field approximations.
	Specifically, this was shown by analyzing the propagation in a Kerr medium.
	Using our technique, we were able to compared and clearly discriminate nonclassicality of the initial states and quantum features which are a result of the quantum evolution itself.

	We then considered semiclassical models in which one part is a quantum system and the other one is classical.
	This is important, for instance, for studying in a consistent manner processes in quantum-classical systems, including interactions between them.
	We demonstrated that our approach allows us to carry out such a task, for example, by considering a classical light field coupled to a two-level atom which led to a classical wave equation for the radiation field and a Schr\"{o}dinger equation for the atom.
	We could show that, despite interacting with the quantum atom for all times, the field remains classical, which cannot be achieved when applying other techniques.

	As a third example of practical relevance for quantum information science, we considered the entanglement dynamics.
	In particular, we analyzed the challenging scenario of a system which consists of a macroscopic number of coupled quantum mechanical oscillators.
	In this scenario, the classical states are identified with separable ones.
	The classical dynamics resulting from our method is described in terms of Schr\"odinger-type equations for each subsystem separately.
	As we have shown for different forms of partial separability (or, conversely, partial entanglement), the entanglement-generating evolution can be clearly distinguished from the classical one, even for initially fully separable mixed states and weakly interacting particles.

	In conclusion, we developed a general and easily accessible method to certify quantum properties over time.
	Our technique is not restricted to a specific type of quantum coherence and applies to arbitrary time scales and interaction regimes.
	It also enables us to separate quantum features within the initial states from those arising from the quantum evolution alone.
	In addition, the evolution of quantum correlations between an arbitrary number of classical and quantum subsystems can be consistently studied and characterized with our method.
	Thus, we believe that our theoretical framework presents a new versatile tool and can be the starting point for future studies of temporal quantum phenomena.

\ack
	The project leading to this application has received funding from the European Union's Horizon 2020 research and innovation program under Grant Agreement No. 665148 (QCUMbER).

\appendix

\section{Hybrid dynamics}\label{app:Hybrid}

	Here we demonstrate in more detail how our technique can be applied.
	Suppose a harmonic oscillator is coupled to a $d$-level system.
	This results in the general Hamiltonian
	\begin{eqnarray}
		\label{eq:HybridH}
		\hat H &=&
		\sum_{j,j'=1}^{d}\hat H_{j,j'}\otimes|j\rangle\langle j'|,
	\end{eqnarray}
	where $\hat H_{j,j'}=\hbar\sum_{k,k'=0}^\infty\Omega_{k,k'}^{(j,j')}\hat a^{\dag k}\hat a^{k'}=\hat H_{j',j}^\dag$ is decomposed in bosonic field operators and using the orthonormal basis $\{|1\rangle,\ldots,|d\rangle\}$.
	In addition, we demand that the state of the composite system is a tensor product of a continuous-variable coherent state and a discrete-variable state,
	\begin{equation}
		\label{eq:HybridS}
		|\psi\rangle=|\alpha\rangle\otimes|\phi\rangle.
	\end{equation}
	Since $|\alpha\rangle=\rme^{-\alpha^\ast\alpha/2}\sum_{n=0}^\infty \alpha^n|n\rangle/\sqrt{n!}$ depends on $\alpha^\ast$, it is convenient to decompose
	\begin{equation}
		\alpha=x_0+\rmi x_1,
	\end{equation}
	using real numbers $x_0$ and $x_1$.
	One could equivalently apply the results of section \ref{subsec:Phase} to the vector $\rme^{|\alpha|^2/2}|\alpha\rangle$, depending only on $\alpha$ but not being normalized.

	For formulating the semiclassical equations of motion \eref{eq:ClSE}, we firstly determine the tangential vectors
	\begin{eqnarray}
		|\partial_{x_m}\psi\rangle &=& (\rmi^m\hat a^\dag-x_m)|\alpha\rangle\otimes|\phi\rangle,
		\\
		|\partial_{\phi_j}\psi\rangle &=& |\alpha\rangle\otimes|j\rangle,
	\end{eqnarray}
	for $m\in\{0,1\}$ and $j\in\{1,\ldots,d\}$.
	Furthermore, we compute the vector
	\begin{eqnarray}
		|\chi\rangle &=& \left[\rmi\hbar\frac{\rmd}{\rmd t}-\hat H\right]|\psi\rangle
		\\ \nonumber
		&=&
		\rmi\hbar\sum_{j}\phi_j\left[
			\dot x_0(\hat a^\dag-x_0)
			+\dot x_1(\rmi\hat a^\dag-x_1)
		\right]|\alpha\rangle\otimes|j\rangle
		\\ \nonumber
		&&
		{+}\rmi\hbar\sum_{j}\dot\phi_j|\alpha\rangle\otimes|j\rangle
		{-}\sum_{j}\Big[\sum_{j'}\phi_{j'}\hat H_{j,j'}\Big]|\alpha\rangle\otimes|j\rangle.
	\end{eqnarray}

	Applying equation \eref{eq:ClSE} to the discrete-variable part, we obtain an evolution described by
	\begin{eqnarray}
		\label{eq:DVEOMprelim}
		0 &=& \langle\partial_{\phi_j}\psi|\chi\rangle
		\\ \nonumber
		&=&
		\hbar[\dot x_0x_1-x_0\dot x_1]\phi_j+\rmi\hbar\dot\phi_j
		-\sum_{j'}\mathcal H_{j,j'}\phi_{j'},
	\end{eqnarray}
	where $\mathcal H_{j,j'}=\langle\alpha|\hat H_{j,j'}|\alpha\rangle$ is a function of $x_0$ and $x_1$.
	The first term in the above equation contributes as a global phase only, thus being negligible.
	In particular, it can be easily transformed to be zero by replacing $|\phi\rangle=\rme^{\rmi \varphi}|\phi'\rangle$, with a time-dependent phase defined through $\dot\varphi=\dot x_0x_1-x_0\dot x_1$, resulting in the Schr\"odinger equation
	\begin{equation}
		\label{eq:DVEOM}
		0=\rmi\hbar\dot\phi'_j-\sum_{j'}\mathcal H_{j,j'}\phi'_{j'},
	\end{equation}
	for the coefficients $\phi'_j$ of the state $|\phi'\rangle$.

	For the continuous-variable subsystem, we apply equation \eref{eq:ClSEreal}, $0 = \mathrm{Re}\left(\langle\partial_{x_m}\psi|\chi\rangle\right)$.
	For this purpose, we firstly compute the expression
	\begin{eqnarray}
		\label{eq:complexCVEOM}
		\langle\partial_{x_m}\psi|\chi\rangle
		&=&
		\rmi\hbar\langle\phi|\phi\rangle\langle\alpha|
			\left[(-\rmi)^m\hat a-x_m\right]
		\\ \nonumber
		&&\times
			\left[\dot \alpha\hat a^\dag-(\dot x_0x_0+\dot x_1x_1)\right]
		|\alpha\rangle
		\\ \nonumber
		&&
		+\rmi\hbar\langle\phi|\dot\phi\rangle\langle\alpha|(-\rmi)^m\hat a-x_m|\alpha\rangle
		\\ \nonumber
		&&
		-\langle\alpha|\left[(-\rmi)^m\hat a-x_m\right]\hat H_\phi|\alpha\rangle,
	\end{eqnarray}
	using $\hat H_\phi=\sum_{j,j'}\phi_j^\ast\phi_{j'}\hat H_{j,j'}=\hbar\sum_{k,k'}\Omega_{k,k'}\hat a^{\dag k}\hat a^{k'}$, with $\Omega_{j,j'}=\sum_{j,j'}\phi_j^\ast\phi_{j'}\Omega_{k,k'}^{(j,j')}$ being a function of the discrete-variable state.
	We now evaluate the individual terms in the above equation.
	Firstly, we have a conserved normalization, $\langle\phi|\phi\rangle=1$.
	Secondly, equation \eref{eq:DVEOMprelim} can be applied to rewrite $\rmi\hbar\langle\phi|\dot\phi\rangle=\mathcal H-\hbar[\dot x_0x_1-x_0\dot x_1]$, where we defined $\mathcal H=\langle\alpha|\hat H_\phi|\alpha\rangle$.
	In addition, we can apply fundamental commutation relations, $[\hat a,\hat a^{\dag k}]=k\hat a^{\dag(k-1)}$, to show $\langle\alpha|\hat a\hat H_\phi|\alpha\rangle=\alpha\mathcal H+\partial_{\alpha^\ast}\mathcal H$ for the last term.
	In this context, recall the decomposition
	\begin{equation}
		\partial_{\alpha^\ast}=\frac{1}{2}\partial_{x_0}+\frac{\rmi}{2}\partial_{x_1}.
	\end{equation}
	Finally, using the above relations and straightforward algebra, we can now rewrite equation \eref{eq:complexCVEOM} as
	\begin{eqnarray}
		&&\frac{\langle\partial_{x_m}\psi|\chi\rangle}{\hbar}
		\\ \nonumber
		&=&
		\rmi(-\rmi)^m(\dot x_0+\rmi\dot x_1)
		-(-\rmi)^m\frac{\partial_{x_0}\mathcal H+\rmi\partial_{x_1}\mathcal H}{2\hbar},
	\end{eqnarray}
	which gives $0 = -\mathrm{Re}\left(\langle\partial_{x_0}\psi|\chi\rangle\right)/\hbar=\dot x_1+\partial_{x_0}\mathcal H/(2\hbar)$
	and $0 = \mathrm{Re}\left(\langle\partial_{x_0}\psi|\chi\rangle\right)/\hbar=\dot x_0-\partial_{x_1}\mathcal H/(2\hbar)$.
	Both equations can be recombined into one result for the complex field amplitude $\alpha=x_0+\rmi x_1$,
	\begin{equation}
		\label{eq:CVEOM}
		0=\dot \alpha+\frac{\rmi}{\hbar}\partial_{\alpha^\ast}\mathcal H.
	\end{equation}

	Summarizing this appendix, we provided a detailed calculation of how our equations of motion apply to a general hybrid Hamilton operator \eref{eq:HybridH} and semiclassical states of the form \eref{eq:HybridS}, resulting in equations \eref{eq:DVEOM} and \eref{eq:CVEOM}.
	It is worth emphasizing that the hybrid dynamics presented in this appendix also yields, as special cases, the results of sections \ref{sec:Sanity} (Schr\"odinger equation), \ref{sec:optics} (nonlinear optics), and \ref{sec:semiclassical} (semiclassical models).

\section{Solution for separable dynamics}\label{app:Entanglement}

\subsection{Notations}

	The Hamilton operator \eref{eq:HmechOsci} of the $N$-partite system under study can be put into the form
	\begin{eqnarray}
		\hat H=\sum_{k=1}^N\left(
			\frac{1}{2m}\hat p_k^2
			{+}\frac{m\omega^2}{2}\hat x_k^2
		\right)
		+\frac{\kappa}{4}\sum_{k,l=1}^N\left(
			\hat x_k{-}\hat x_l
		\right)^2,
	\end{eqnarray}
	where $[\hat x_k,\hat p_l]=\rmi\hbar\delta_{k,l}$, with $\delta_{k,l}$ denoting the Kronecker symbol.
	In particular, we can identify the specific form of canonical operators in position representation as $\vec{\hat x}=(x_1,\ldots,x_N)^\mathrm{T}$ and $\vec{\hat p}=(\hbar/\rmi)(\partial/\partial x_1,\ldots,\partial/\partial x_N)^\mathrm{T}$.
	Following the approach in \cite{SW17}, it is convenient to consider the system's dynamics in ``natural'' units, leading to the following quantities:
	a rescaled time, $\tau=\omega t$;
	a coupling constant, $R=\kappa/(m\omega^2)$;
	transformed canonical coordinates, $\hat \xi_k=(m\omega/\hbar)^{1/2}x_k$ and $\hat \pi_k=-\rmi\partial/\partial\xi_k$;
	as well as a rescaled Hamilton operator,
	\begin{eqnarray}
		\hat \eta=\frac{\hat H}{\hbar\omega}
		=\frac{1}{2}\vec{\hat\pi}^\mathrm{T}\vec{\hat\pi}
		+\frac{1+RN}{2}\vec{\hat\xi}^\mathrm{T}\vec{\hat\xi}
		-\frac{R}{2}\vec{\hat\xi}^\mathrm{T}\vec n\vec n^\mathrm{T}\vec{\hat\xi},
	\end{eqnarray}
	where $[\hat \xi_k,\hat\pi_l]=\rmi\delta_{k,l}$ and $\vec n=(1,\ldots,1)^\mathrm{T}$.
	Note that for $R=0$, we obtain the noninteracting case.

	We aim to study the time-dependent behavior of second-order correlations.
	Because of the noncommuting property of position and momentum, let us briefly mention the noncommuting components of the covariance matrix $C$, $\langle \hat\xi_k\hat\pi_l+\hat\pi_l\hat\xi_k\rangle/2-\langle \hat\xi_k\rangle\langle\hat\pi_l\rangle$.
	We further have $\hat\xi_k\hat\pi_l=(\rmi\delta_{k,l}+\hat\xi_k\hat\pi_l+\hat\pi_l\hat\xi_k)/2$ and $\hat\pi_l\hat\xi_k=(-\rmi\delta_{k,l}+\hat\xi_k\hat\pi_l+\hat\pi_l\hat\xi_k)/2$, which yields
	\begin{eqnarray}
		\nonumber
		\left(\begin{array}{c}\vec{\hat\xi}\\\vec{\hat\pi}\end{array}\right)
		\left(\begin{array}{c}\vec{\hat\xi}\\\vec{\hat\pi}\end{array}\right)^\mathrm{T}
		&=&\left(\begin{array}{cc}
			\frac{\vec{\hat\xi}\vec{\hat\xi}^\mathrm{T}+[\vec{\hat\xi}\vec{\hat\xi}^\mathrm{T}]^\mathrm{T}}{2}
			& \frac{\vec{\hat\xi}\vec{\hat\pi}^\mathrm{T}+[\vec{\hat\xi}\vec{\hat\pi}^\mathrm{T}]^\mathrm{T}}{2}
			\\
			\frac{\vec{\hat\pi}\vec{\hat\xi}^\mathrm{T}+[\vec{\hat\pi}\vec{\hat\xi}^\mathrm{T}]^\mathrm{T}}{2}
			& \frac{\vec{\hat\pi}\vec{\hat\pi}^\mathrm{T}+[\vec{\hat\pi}\vec{\hat\pi}^\mathrm{T}]^\mathrm{T}}{2}
		\end{array}\right)
		\\
		&&+\frac{\rmi}{2}
		\underbrace{\left(\begin{array}{cc}
			0 & E \\ -E & 0
		\end{array}\right)}_{=J},
	\end{eqnarray}
	where $E=\mathrm{diag}(1,\ldots,1)$ is the $N\times N$ identity and $J=-J^\mathrm{T}$ is the symplectic matrix.
	It is also worth recalling that expectation values of $J$ with real-valued vectors are zero.

\subsection{Quantum dynamics}

	As a consequence of the above considerations, the full quantum dynamics is described by the following equations of motion: $\rmd\hat\xi_k/\rmd\tau=-\rmi[\hat\xi_k,\hat\eta]$ and $\rmd\hat\pi_k/\rmd\tau=-\rmi[\hat\pi_k,\hat\eta]$.
	In addition, we have
	\begin{eqnarray}
		&&\frac{\rmd}{\rmd\tau}\left(\begin{array}{c}\vec{\hat\xi}\\\vec{\hat\pi}\end{array}\right)
		\left(\begin{array}{c}\vec{\hat\xi}\\\vec{\hat\pi}\end{array}\right)^\mathrm{T}
		\\\nonumber
		&=&
		\left[\frac{\rmd}{\rmd\tau}\left(\begin{array}{c}\vec{\hat\xi}\\\vec{\hat\pi}\end{array}\right)\right]
		\left(\begin{array}{c}\vec{\hat\xi}\\\vec{\hat\pi}\end{array}\right)^\mathrm{T}
		+\left(\begin{array}{c}\vec{\hat\xi}\\\vec{\hat\pi}\end{array}\right)
		\left[\frac{\rmd}{\rmd\tau}\left(\begin{array}{c}\vec{\hat\xi}\\\vec{\hat\pi}\end{array}\right)\right]^\mathrm{T}.
	\end{eqnarray}
	Because it is also useful for the separable case, let us assume for the time being that the Hamilton operator takes the more general form
	\begin{eqnarray}
		\hat\eta=\frac{1}{2}\vec{\hat\pi}^\mathrm{T}\vec{\hat\pi}+\frac{1}{2}\vec{\hat\xi}^\mathrm{T}G\vec{\hat\xi}+\vec{v}^\mathrm{T}\vec{\hat\xi},
	\end{eqnarray}
	with an $N$-dimensional vector $\vec v$ and a real-valued, symmetric, and positive definite $N\times N$ matrix $G$.
	Then we get
	\begin{eqnarray}
		\frac{\rmd}{\rmd\tau}\left(\begin{array}{c}\vec{\hat\xi}\\\vec{\hat\pi}\end{array}\right)
		=J
		\left(\begin{array}{cc} G & 0 \\ 0 & E \end{array}\right)
		\left(\begin{array}{c}\vec{\hat\xi}\\\vec{\hat\pi}\end{array}\right)
		+\left(\begin{array}{c}\vec v\\0\end{array}\right).
	\end{eqnarray}

	Furthermore, using the definition $\Delta\hat y=\hat y-\langle\hat y\rangle$ for any $\hat y$, we get
	\begin{eqnarray}
		\nonumber
		&&\frac{\rmd}{\rmd\tau}
		\left(\begin{array}{cc}
			\langle\Delta\vec{\hat\xi}\Delta\vec{\hat\xi}^\mathrm{T}\rangle
			& \langle\Delta\vec{\hat\xi}\Delta\vec{\hat\pi}^\mathrm{T}\rangle
			\\
			\langle\Delta\vec{\hat\pi}\Delta\vec{\hat\xi}^\mathrm{T}\rangle
			& \langle\Delta\vec{\hat\pi}\Delta\vec{\hat\pi}^\mathrm{T}\rangle
		\end{array}\right)
		\\\nonumber
		&=&J\left(\begin{array}{cc} G & 0 \\ 0 & E \end{array}\right)
		\left(\begin{array}{cc}
			\langle\Delta\vec{\hat\xi}\Delta\vec{\hat\xi}^\mathrm{T}\rangle
			& \langle\Delta\vec{\hat\xi}\Delta\vec{\hat\pi}^\mathrm{T}\rangle
			\\
			\langle\Delta\vec{\hat\pi}\Delta\vec{\hat\xi}^\mathrm{T}\rangle
			& \langle\Delta\vec{\hat\pi}\Delta\vec{\hat\pi}^\mathrm{T}\rangle
		\end{array}\right)
		\\
		&&-\left(\begin{array}{cc}
			\langle\Delta\vec{\hat\xi}\Delta\vec{\hat\xi}^\mathrm{T}\rangle
			& \langle\Delta\vec{\hat\xi}\Delta\vec{\hat\pi}^\mathrm{T}\rangle
			\\
			\langle\Delta\vec{\hat\pi}\Delta\vec{\hat\xi}^\mathrm{T}\rangle
			& \langle\Delta\vec{\hat\pi}\Delta\vec{\hat\pi}^\mathrm{T}\rangle
		\end{array}\right)
		\left(\begin{array}{cc} G & 0 \\ 0 & E \end{array}\right)J,
	\end{eqnarray}
	which is independent of $\vec v$.
	The solution is described in terms of the symplectic transformation
	\begin{eqnarray}
		\nonumber
		&&S(\tau)
		\\
		&=&\left(\begin{array}{cc}
			\cos(G^{1/2}\tau)
			&
			G^{-1/2}\sin(G^{1/2}\tau)
			\\
			-G^{1/2}\sin(G^{1/2}\tau)
			&
			\cos(G^{1/2}\tau)
		\end{array}\right),
	\end{eqnarray}
	which satisfies
	\begin{eqnarray}
		S(0)&=&\left(\begin{array}{cc} E & 0 \\ 0 & E \end{array}\right),
		\\
		\frac{\rmd S(\tau)}{\rmd\tau}&=&J\left(\begin{array}{cc} G & 0 \\ 0 & E \end{array}\right)S(\tau).
	\end{eqnarray}
	This symplectic transformation applies to an initial covariance matrix $C(0)$ as $C(\tau)=S(\tau)C(0)S(\tau)^\mathrm{T}$.
	As a result of the spectral decomposition of the specific case $G=\mu E+\nu\vec n\vec n^\mathrm{T}$, we get
	\begin{eqnarray}
		\nonumber
		G^{1/2}
		&=&\sqrt{\mu}\left(E-\left[\frac{\vec n}{\sqrt{N}}\right]\left[\frac{\vec n}{\sqrt{N}}\right]^\mathrm{T}\right)
		\\
		&&+\sqrt{\mu+\nu N}\left[\frac{\vec n}{\sqrt{N}}\right]\left[\frac{\vec n}{\sqrt{N}}\right]^\mathrm{T},
	\end{eqnarray}
	for the $N$-partite system under study.

\subsection{Separable dynamics}

	For studying separable dynamics (see \cite{SW17a} and its supplemental material for an detailed introduction), let us decompose the $N$-partite system into $K$ subsystems, each consisting of $N_j$ individual oscillators, for $j=1,\ldots,K$.
	In this scenario, it is useful to decompose, for example, the generalized position as
	\begin{equation}
		\vec{\hat\xi}=\left(\begin{array}{c}
			\vec{\hat\xi}_1
			\\\vdots\\
			\vec{\hat\xi}_K
		\end{array}\right)
	\end{equation}
	and the separable state as $|\psi\rangle=|\psi_1\rangle\otimes\cdots\otimes|\psi_K\rangle$.
	The partially reduced Hamilton operator for the $j$th subsystem takes the form
	\begin{eqnarray}
		\nonumber
		&&\hat \eta_{\psi_{1},\ldots,\psi_{j-1},\psi_{j+1},\ldots,\psi_{K}}
		\\\nonumber
		&=&\frac{1}{2}\vec{\hat\pi}_j^\mathrm{T}\vec{\hat\pi}_j
		+\frac{1+RN}{2}\vec{\hat\xi}_j^\mathrm{T}\vec{\hat\xi}_j
		\\
		&&-\frac{R}{2}\vec{\hat\xi}_j^\mathrm{T}\vec n_j\vec n_j^\mathrm{T}\vec{\hat\xi}_j
		-R\left(\sum_{l:l\neq j}\vec n_l^\mathrm{T}\langle\vec{\hat\xi}_l\rangle_{\psi_l}\right)\vec n_j^\mathrm{T}\vec{\hat\xi}_j,
	\end{eqnarray}
	where contributions proportional to $\hat 1_j$ have been ignored as they only contribute to the propagation in time with a global phase.

	Applying our previous results, we readily find for the expectation values of $\vec{\hat\xi}_k$ and $\vec{\hat\pi}_K$ equations of motion which are identical to the entangled case.
	However, the covariance of the $j$th component evolves in the separable scenario differently---namely, according to
	\begin{eqnarray}
		&&\frac{\rmd}{\rmd\tau}
		\left(\begin{array}{cc}
			\langle\Delta\vec{\hat\xi}_j\Delta\vec{\hat\xi}_j^\mathrm{T}\rangle_{\psi_j}
			& \langle\Delta\vec{\hat\xi}_j\Delta\vec{\hat\pi}_j^\mathrm{T}\rangle_{\psi_j}
			\\
			\langle\Delta\vec{\hat\pi}_j\Delta\vec{\hat\xi}_j^\mathrm{T}\rangle_{\psi_j}
			& \langle\Delta\vec{\hat\pi}_j\Delta\vec{\hat\pi}_j^\mathrm{T}\rangle_{\psi_j}
		\end{array}\right)
		\\\nonumber
		&=&J_j\left(\!\!\begin{array}{cc} G_j & 0 \\ 0 & E_j \end{array}\!\!\right)\!\!
		\left(\!\!\begin{array}{cc}
			\langle\Delta\vec{\hat\xi}_j\Delta\vec{\hat\xi}_j^\mathrm{T}\rangle_{\psi_j}
			& \langle\Delta\vec{\hat\xi}_j\Delta\vec{\hat\pi}_j^\mathrm{T}\rangle_{\psi_j}
			\\
			\langle\Delta\vec{\hat\pi}_j\Delta\vec{\hat\xi}_j^\mathrm{T}\rangle_{\psi_j}
			& \langle\Delta\vec{\hat\pi}_j\Delta\vec{\hat\pi}_j^\mathrm{T}\rangle_{\psi_j}
		\end{array}\!\!\right)
		\\\nonumber
		&&-\left(\!\!\begin{array}{cc}
			\langle\Delta\vec{\hat\xi}_j\Delta\vec{\hat\xi}_j^\mathrm{T}\rangle_{\psi_j}
			& \langle\Delta\vec{\hat\xi}_j\Delta\vec{\hat\pi}_j^\mathrm{T}\rangle_{\psi_j}
			\\
			\langle\Delta\vec{\hat\pi}_j\Delta\vec{\hat\xi}_j^\mathrm{T}\rangle_{\psi_j}
			& \langle\Delta\vec{\hat\pi}_j\Delta\vec{\hat\pi}_j^\mathrm{T}\rangle_{\psi_j}
		\end{array}\!\!\right)\!\!
		\left(\!\!\begin{array}{cc} G_j & 0 \\ 0 & E_j \end{array}\!\!\right)J_j,
	\end{eqnarray}
	with $G_j=(1+RN)E_j-R\vec n_j\vec n_j^\mathrm{T}$.
	Note that quantities previously defined without an index are similarly defined for the $j$th subspace.
	Let us emphasize that, because of factorization, we have $\langle\Delta \hat y\otimes\Delta\hat z\rangle_{\psi\otimes\phi}=\langle\Delta \hat y\rangle_{\psi}\langle\Delta\hat z\rangle_{\phi}=0$ for all $\hat y$ and $\hat z$ and all times.
	Thus, the evolution of the covariance in the separable case is fully described by the evolution of the $j$th block of the corresponding subsystem separately, which is given by
	\begin{eqnarray}
		\nonumber
		&&S_j(\tau)
		\\
		&=&\left(\begin{array}{cc}
			\cos(G_j^{1/2}\tau)
			&
			G^{-1/2}\sin(G_j^{1/2}\tau)
			\\
			-G^{1/2}\sin(G_j^{1/2}\tau)
			&
			\cos(G_j^{1/2}\tau)
		\end{array}\right),
	\end{eqnarray}
	with
	\begin{eqnarray}
		\nonumber
		G_j^{1/2}
		&=&\sqrt{1+RN}\left(E_j-\left[\frac{\vec n_j}{\sqrt{N_j}}\right]\left[\frac{\vec n_j}{\sqrt{N_j}}\right]^\mathrm{T}\right)
		\\
		&&+\sqrt{1+R(N{-}N_j)}\left[\frac{\vec n_j}{\sqrt{N_j}}\right]\left[\frac{\vec n_j}{\sqrt{N_j}}\right]^\mathrm{T}.
	\end{eqnarray}

\subsection{Initial values}

	Finally, the specific initial condition is the thermal state of the noninteracting system, $R=0$.
	In this case, the covariance decomposes in one-partite blocks with the Hamiltonian $\hat \eta_j=(\hat\pi_j^2+\hat\xi_j^2)/2$.
	The partition function reads $Z_j=\mathrm{tr}(\rme^{-\beta\hat\eta_j})=\rme^{-\beta/2}/(1-\rme^{-\beta})$, where $\beta=\hbar\omega/(k_B T)$ in our natural units.
	Further, this yields the desired moments, $\langle\hat\xi_j\rangle=\langle\hat\pi_j\rangle=\langle \hat\xi_j\hat\pi_j+\hat\pi_j\hat\xi_j\rangle/2=0$ and $\langle\hat\xi_j^2\rangle=\langle\hat\pi_j^2\rangle=1/2+e^{-\beta}/(1-e^{-\beta})$.
	Thus, the $j$th block of the propagated covariance matrix, $C_j(\tau)$, reads
	\begin{eqnarray}
		&&C_j(\tau)
		\\\nonumber
		&=&\left[\frac{1}{2}+\frac{\rme^{-\beta}}{1-\rme^{-\beta}}\right]\cos^2(G_j^{1/2}\tau)
		\left(\begin{array}{cc}
			E_j & 0
			\\
			0 & E_j
		\end{array}\right)
		\\\nonumber
		&&+\left[\frac{1}{2}+\frac{\rme^{-\beta}}{1-\rme^{-\beta}}\right]\frac{\sin^2(G_j^{1/2}\tau)}{2}
		\left(\begin{array}{cc}
			G_j^{-1} & 0
			\\
			0 & G_j
		\end{array}\right).
	\end{eqnarray}
	For the variance of the momentum operator parallel to $\vec n$, \textit{i.e.} $\hat\Pi=\vec n^\mathrm{T}\vec{\hat\pi}/N$, we simply find
	\begin{eqnarray}
		&&V_\Pi(\tau)
		=\frac{1}{N^2}
		\left(\begin{array}{c}0\\\vec n\end{array}\right)^\mathrm{T}
		C(\tau)\left(\begin{array}{c}\vec 0\\\vec n\end{array}\right)
		\\\nonumber
		&=& \frac{1}{N}\left[\frac{1}{2}+\frac{\rme^{-\beta}}{1-\rme^{-\beta}}\right][1+r_K\sin^2(\sqrt{1+r_K}\,\tau)],
	\end{eqnarray}
	using $N_1=\cdots=N_K=N/K$ and $r_K=RN(1-1/K)$.


\end{document}